\documentclass[floats,floatfix,showpacs,amssymb,prd,twocolumn,superscriptaddress,nofootinbib,nolongbibliography,reprint]{revtex4-2}

\usepackage{amssymb,amsmath,verbatim,mathtools,needspace,enumitem,etoolbox,graphicx,physics,microtype,afterpage,bigints,gensymb,tabularx}

\usepackage[dvipsnames, usenames]{xcolor}
\definecolor{linkcolor}{rgb}{0.0,0.3,0.5}
\definecolor{dodgerblue}{HTML}{1E90FF}
\usepackage[unicode, colorlinks=true, linkcolor=linkcolor, citecolor=linkcolor, filecolor=linkcolor,urlcolor=linkcolor, pdfusetitle]{hyperref}
\usepackage[all]{hypcap}
\usepackage[T1]{fontenc}
\usepackage[utf8]{inputenc}
\usepackage{orcidlink}
\usepackage{bm}
\usepackage{empheq} 

\newcommand{\ssim}{\mathchar"5218\relax\,}

\makeatletter
\newcommand*{\balancecolsandclearpage}{\close@column@grid \cleardoublepage \twocolumngrid}
\makeatother

\usepackage{lipsum}

\def\araa{{Ann.~Rev.~Astron.~Astrophys.}}
\def\apj{{Astrophys.~J.}}\def\apjl{{Astrophys.~J.~Lett.}}

\def\mnras{{Mon.~Not.~R.~Astron.~Soc.}}

\def\prd{{Phys.~Rev.~D}}

\def\prl{{Phys.~Rev.~Lett.}}

\def\nat{{Nature}}

\def\cqg{{Class. Quantum Gravity}}
\def\natastro{{Nat. Astron.}}

\newcommand{\bham}{\affiliation{School of Physics and Astronomy \& Institute for Gravitational Wave Astronomy, University of Birmingham, \\ Birmingham, B15 2TT, United Kingdom}}

\newcommand{\milan}{\affiliation{Dipartimento di Fisica ``G. Occhialini'', Universit\'a degli Studi di Milano-Bicocca, Piazza della Scienza 3, 20126 Milano, Italy}}
\newcommand{\infn}{\affiliation{INFN, Sezione di Milano-Bicocca, Piazza della Scienza 3, 20126 Milano, Italy}}

\begin{document}

\title{Spin-eccentricity interplay in merging binary black holes}

\author{Giulia Fumagalli$\,$\orcidlink{0009-0004-2044-989X}}
\email{g.fumagalli47@campus.unimib.it}
\milan \infn 

\author{Davide Gerosa$\,$\orcidlink{0000-0002-0933-3579}}

\milan \infn \bham


\date{\today}

\begin{abstract}

Orbital eccentricity and spin precession are precious observables to infer the formation history of binary black holes with gravitational-wave data. 
We present a post-Newtonian, multi-timescale analysis of the binary dynamics able to capture both precession and eccentricity over long inspirals. 
We show that the evolution of an eccentric binary can be reduced that of effective source on quasi-circular orbits, coupled to a post-Newtonian prescription for the secular evolution of the eccentricity. 
Our findings unveil an interplay between precession and eccentricity: the spins of eccentric binaries precess on shorter timescales and their nutation amplitude is altered compared to black holes on quasi-circular orbits, consequently affecting the so-called spin morphology. 
Even if binaries circularize by the time they enter the sensitivity window of our detectors, their spin orientations retain some memory of the past evolution on eccentric orbits, thus providing a new link between gravitational-wave detection and astrophysical formation. %
 At the same time, we point out that residual eccentricity should be considered a source of systematics when reconstructing the past history of black-hole binaries using the spin orientations.

\end{abstract}

\maketitle

\section{Introduction}

\label{secintro}

The most generic bound orbits in Kepler's two-body problem are ellipses, which are parametrized by their eccentricity $0\leq e <1$. In General Relativity, point particles are substituted by binary black holes (BHs) and, at least in the post-Newtonian (PN) regime, the underlying timescale separation implies that orbits can still be treated using conic sections~\cite{1964PhRv..136.1224P}. %
Gravitational-wave (GW) emission causes a secular evolution of the orbital parameters that tends to both shrink and circularize the orbit. The relativistic two-body problem also presents distinctive features related to the BH spins. Binaries with spins that are misaligned with respect to the orbital angular momentum precess about the direction of the total angular momentum, varying their orientation%
~\cite{1994PhRvD..49.6274A}.

Both eccentricity and spin precession are expected to leave an imprint
on the emitted GW signals %
and their simultaneous measurement can provide crucial hints on the undelying formation and evolutionary processes. %
For stellar-mass BHs, orbital eccentricity is predicted to be a clean signature of BH binaries assembled via recent dynamical interactions~\cite{2018PhRvD..97j3014S,2021ApJ...907L..20T,2021ApJ...921L..43Z}, notably including Kozai-Lidov oscillations in triple systems \cite{2016ARA&A..54..441N}. %
Misaligned spins are also expected to be a telltale of GW sources formed dynamically~\cite{2016ApJ...832L...2R,2017Natur.548..426F,2017NatCo...814906S,2021NatAs...5..749G}, though this is probably not a unique feature~\cite{2013PhRvD..87j4028G,2018PhRvD..98h4036G,2021PhRvD.103f3032S}. 
For supermassive BH binaries, eccentricity and spins both encode information on the %
 pairing and hardening processes that lead to the formation of such systems following galaxy mergers~\cite{2021FrASS...8....7S}.

At present, LIGO/Virgo data shows tantalizing evidence of both eccentricity \cite{2022ApJ...940..171R} and spin precession \cite{2022Natur.610..652H,2022PhRvD.106j4017P} in a few binary BH events. The combined inference of the two effects is still beyond the horizon because of the related waveform-modeling challenges \cite{2021PhRvD.103f4022I,2022PhRvD.105d4035R, 2023arXiv230409662N,2022ApJ...936..172K}
as well as  potential degeneracies~\cite{2023MNRAS.519.5352R}.

LIGO and Virgo can distinguish eccentricities $e \gtrsim 0.05$, future ground third-generation detectors will push this limit to $e \gtrsim 10^{-4}$~\cite{2018PhRvD..98h3028L}, while LISA will be sensible to $e \gtrsim 10^{-2.5}$ for supermassive BH binaries~\cite{2023arXiv230713367G}. 
In practice, sources with eccentricities smaller than these thresholds are to be considered circular for observational purposes.

For context, an equal-mass BHs binary with total mass $M=20$ $M_{\odot}$ evolving in isolation under radiation reaction from an initial
semi-major axis %
$a= 30$ $R_{\odot}$ and an initial eccentricity $e=0.6$ will reach the innermost stable orbit with eccentricity $e \sim 10^{-4}$. %
 In other words, PN circularization is brutal, and most binaries that are formed on highly eccentric orbits might enter the detector sensitivity range with eccentricity values that are well below the distinguishability threshold \cite{2018PhRvD..98h3028L,2021ApJ...921L..43Z}. While hoping for prominent outliers \cite{2018PhRvD..97j3014S}, solid inference on orbital eccentricity might well remain challenging. %

BH binaries that are observed as circular might have spent a large fraction of their lifetime on eccentric orbits. Is this information somewhat accessible to GW detectors? Do merging BHs retain any memory of their eccentric past? In this paper, we explore this line of reasoning and show some information is indeed encoded in the spin sector of the gravitational dynamics. The evolution of the BH spins depends on the binary eccentricity \cite{2010PhRvD..81l4001K,2018PhRvD..98j4043K,2019PhRvD.100l4008P,2020PhRvD.102l3009Y,2021arXiv210610291K} opening for the exciting possibility of using accurate spin measurements to infer the the presence of eccentricity at BH formation, even if this does not correspond to eccentricity at detection. In particular, this paper focuses on the methodological aspects of this problem and develops the appropriate formalism to capture long %
evolutions of precessing, eccentric BH binaries.

Specifically, we exploit a multi-timescale approximation to the inspiral of BH binaries, which relies on averaging the dynamics in sequence over the short timescales of the problem (orbit and spin precession) while introducing the evolution over the longer timescale (radiation reaction) in a quasi-adiabatic fashion.
This framework has been developed~\cite{2015PhRvL.114h1103K,2015PhRvD..92f4016G} and extensively tested~\cite{2022PhRvD.106b3001J,2023PhRvD.108b4042G} for circular binaries and is here fully generalized to eccentric sources (see also Refs.~\cite{2019PhRvD.100l4008P,2020PhRvD.102l3009Y,2021arXiv210610291K} for other attempts in this direction).

In a nutshell, the inclusion of eccentricity requires two ingredients (Sec.~\ref{sec:Eccentric and precessing BBH evolution}): 
\begin{itemize}
\item[(i)] First, we show that the short-timescale dynamics of any eccentric binary can be mapped to that of an effective circular source. This is an extremely elegant property of the gravitational-two body problem and allows us to make direct use of the existing numerical infrastructure \cite{2023PhRvD.108b4042G}. %

\item[(ii)] One then needs to add an evolutionary prescription for the decay of the orbital eccentricity on the radiation-reaction timescale. At leading order, this is the seminal result by \citeauthor{1964PhRv..136.1224P}~\cite{1964PhRv..136.1224P}, which is here presented in a regularized form that allows for safe numerical evaluations at arbitrarily low eccentricities.
\end{itemize}
We then scope out some of the predictions of our new formalism (Sec.~\ref{appsection}). In particular, we investigate the robustness of the underlying timescale hierarchy and the impact of eccentricity on the evolution of precessing BH binaries. Our results are framed in terms of potential GW observables such as the tilt angles and the spin morphologies. 
The investigated couplings between spins and eccentricity provide a new phenomenological handle to infer the past history of BH binaries. The other side of the same coin, however, is that residual eccentricity, if neglected, might introduce a significant systematics in our astrophysical inference (Sec.~\ref{sec: Conclusion}). 
We use natural units where $G = c = 1$.

\section{Binary evolution}\label{sec:Eccentric and precessing BBH evolution}

{We organize our discussion according to the three phenomena that characterize the problem: orbit, precession, and radiation reaction. We then present a timescale comparison.}

\subsection{Orbit}\label{sec:Orbital equations}
\begin{figure}[t]\label{fig:ellipse}
   \includegraphics[width=\columnwidth]{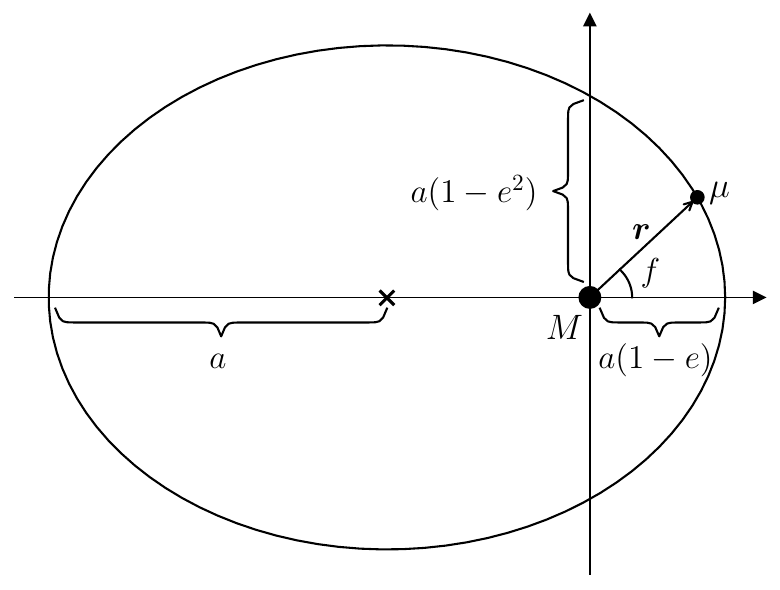} 
    \caption{Binaries on elliptic orbits can be reduced to an effective system with fixed mass $M$ located in one of the foci and an orbiting body of mass $\mu$ (black circles). The geometry of the orbit is defined by semi-major axis $a$, semi-latus rectum $a\left(1-e^2\right)$, and distance at periapsis $a(1-e)$. The true anomaly $f$ is measured counterclockwise from the periapsis to the orbital separation vector $\bm{r}$.} 
\end{figure}

Let us consider a BH binary with component masses $m_{1,2}$, total mass $M=m_1+m_2$, reduced mass is $\mu=(m_1 m_2)/M$, and mass ratio $q=m_2/m_1\leq 1$. In the center-of-mass frame, the problem reduces to an effective scenario where a single particle of mass $\mu$ experiences a central acceleration $\dd^2 \bm{r}/\dd t^2=-M\boldsymbol{r} /r^3$, where $\boldsymbol{r}$ is separation between the two bodies. A schematic representation is provided in Fig.~\ref{fig:ellipse}.

The specific energy and orbital angular momentum are given by 
\begin{align}
   E&=\frac{1}{2}\mu \left(\frac{\dd r}{\dd t}\right)^2+\frac{L^2}{2 \mu r^2}-\frac{\mu M}{r}\,, \\
   L&=\mu r^2 \frac{\dd f}{\dd t}\,,
\end{align}
where $f$ is the true anomaly,
i.e. the angular separation along the orbit measured from the periapsis. 
In Newtonian gravity, orbital energy and angular momentum are conserved and, %
fully characterize the semi-major axis $a$ and the eccentricity $e$: 
\begin{align}
\label{e2}
|E| &= \frac{M q }{2 (1+q) a},
\\
\label{l2}
    L&=\frac{q^2}{(1+q)^2}\sqrt{M^3 a \left(1-e^2\right)}\,.
\end{align}

Elliptical orbits correspond to $e\in (0,1)$, while the limits $e=0$ and $e=1$ corresponds to circles and parabolae, respectively. The orbital separation is given by 
\begin{equation} 
r=\frac{a(1 - e^2)}{1+e\cos f}, 
\end{equation} 
the semi-latus rectum is given by $a\left(1-e^2\right)$, and closest distance between the two bodies is given by $a(1-e)$. From Kepler's third law, the period is
\begin{equation}\label{eq:tau_orb}
    \tau_{\rm orb}=2 \pi \sqrt{\frac{a^3}{M}}\,,
\end{equation} 
which defines the typical timescale of the orbital motion.

Equations (\ref{e2}) and (\ref{l2}) imply that binaries with the same semi-major axis but different eccentricities have the same energy but different angular momentum. In particular, $L$ is smaller for elliptic orbits compared to circular orbits. As explored at length below, this turns out to be a crucial feature when considering spinning BHs. 

Throughout this paper, we refer to ``eccentricity'' as given by its Keplerian definition in terms of $E$ and $L$. This is only appropriate as long as one can approximate the binary evolution as an adiabatic series of quasi-closed orbits, which is typically true in the PN regime for moderate eccentricities (cf. Sec.~\ref{sec: Conclusion}). In contrast, the full theory of GR does not allow for a natural definition of eccentricity, which led to various proposals (see Refs.~\cite{2019CQGra..36b5004L,2023arXiv230211257A} and references therein). %

 \subsection{Spin precession}\label{sec:Precession equations}
A spinning BH binary is characterized by three angular momenta: the spins $\bm{S}_{1,2}$ of the two compact objects and the orbital angular momentum $\bm{L}$. These add up to the total angular momentum $\bm{J}=\bm{L}+\bm{S}_1+\bm{S}_2$. 
The magnitudes of the spins are most commonly quoted in terms of the dimensionless Kerr parameters $\chi_{1,2} = S_{1,2} / m_{1,2}^2 \in [0,1]$.

 We describe the spin orientations using the polar angles $\theta_{1,2} \in [0,\pi]$ between each of the spin vectors and the angular momentum and the azimuthal angle $\Delta \Phi \in [-\pi,\pi]$ between the projections of the two spins onto the orbital plane. 

These quantities can be used to construct the effective spin~\cite{2001PhRvD..64l4013D}
\begin{align}
\chi_{\rm eff}=\frac{\chi_1 \cos\theta_1 +q \chi_2\cos \theta_2}{1+q},
\end{align}
and the weighted spin difference~\cite{2021arXiv210610291K}
\begin{equation}
\delta \chi=\frac{\chi_1 \cos \theta_1 -q \chi_2\cos \theta_2}{1+q}.
\end{equation}
Notably, $\chi_{\rm eff}$ is a constant of motion at 2PN \cite{2008PhRvD..78d4021R} while $\delta \chi$ varies on the precession timescale \cite{2023PhRvD.108b4042G}. 

Relativistic couplings between the orbital angular momentum and the spins cause them to precess about the direction of $\boldsymbol{J}$~\cite{1994PhRvD..49.6274A}. This is a secular motion that takes place over many orbits, which implies the spin dynamics can be consistently orbit-averaged~\cite{2004PhRvD..70l4020S}.
The resulting evolutionary equations at 2PN are \cite{2004PhRvD..70l4020S, 2008PhRvD..78d4021R,2021arXiv210610291K}
\begin{align}
\label{eq:spin1}
\frac{\dd \bm{S}_1}{\dd t}&=\bm{\omega}_1 \times \bm{S}_1,\\
\label{eq:spin2}
\frac{\dd \bm{S}_2}{\dd t}&=\bm{\omega}_2 \times \bm{S}_2,\\ 
\label{eq:dldt}
\frac{\dd \bm{L}}{\dd t}&=\bm{\omega}_L \times \bm{L} +\frac{\dd L}{\dd t}\hat{\bm{L}},
\end{align} 
where 
\begin{align}
\label{eq:spinomega1}
\bm{\omega}_1 &= \frac{1}{2 a^3 \left(1\!-\!e^2\right)^{3/2}}\! \left\{ \left[ 4\!+\!3q \! -\! \frac{3 (1\!+\!q)\chi_{\rm eff }}{\sqrt{a \left(1\!-\!e^2\right) \!/M}}\right] \! \bm{L} \!+\! \bm{S}_2\right\} \!,
\\
\label{eq:spinomega2}
\bm{\omega}_2 &= \frac{1}{2 a^3 \left(1\!-\!e^2\right)^{3/2}}\!\left\{ \left[ 4\!+\!\frac{3}{q}  \!-\! \frac{3 (1\!+\!q)\chi_{\rm eff}}{q\sqrt{a \left(1\!-\!e^2\right)\!/ M}}\right]\!  \bm{L} \!+\! \bm{S}_1\right\} \!,
\\ \nonumber
\bm{\omega}_L &= \frac{1}{2 a^3 \left(1\!-\!e^2\right)^{3/2}}\Bigg\{ \! \left[ 4\!+\!3q \! -\! \frac{3 (1\!+\!q)\chi_{\rm eff }}{\sqrt{a \left(1\!-\!e^2\right) \!/M}}\right] \! \bm{S}_1 
\\ &\qquad\qquad\qquad\quad\,+\left[ 4\!+\!\frac{3}{q}  \!-\! \frac{3 (1\!+\!q)\chi_{\rm eff}}{q\sqrt{a \left(1\!-\!e^2\right)\!/ M}}\right]\! \bm{S}_2\Bigg\},
\label{eq:OmegaL}
\end{align}
model the conservative dynamics and $\dd L/ \dd t$ encodes dissipation via GWs. 

Neglecting radiation reaction and exploiting the conservation of $\chi_{\rm eff}$, one can reduce Eqs.~(\ref{eq:spin1})-(\ref{eq:OmegaL}) to a single equation for the weighted spin difference \cite{2023PhRvD.108b4042G} 
\begin{align}
\label{eq:ddeltachidt}
M \frac{\dd\delta \chi }{\dd t} &= \frac{3 q}{(1+q)^2}  \chi_1\chi_2  \left(1-e^2\right)^{3/2}  \left[\frac{a\left(1-e^2\right)}{M}\right]^{-3} \notag \\
&\times \left[ 1 - \frac{\chi_{\rm eff}}{\sqrt{a\left(1-e^2\right)/M}}\right]
  \sin\theta_1 \sin\theta_2 \sin\Delta\Phi\,,
\end{align}
where the angles $\theta_1$, $\theta_2$, and $\Delta\Phi$ depend on $\delta\chi$.
The solution is quasi-periodic, with $\delta \chi$ oscillating between two extrema $\delta \chi_\pm$ which are themselves functions of $a$, $e$, $q$, $\chi_1$, $\chi_2$, $\chi_{\rm eff}$, and $J$.
In turn, these are used to define the spin precession period
\begin{equation}\label{eq:tauPrec}
\tau_{\rm pre}=2\int_{\delta\chi_-}^{\delta\chi_+}\left(\frac{\dd \delta\chi}{\dd t }\right)^{-1} \dd\delta\chi\,,
\end{equation}
which can be expressed  in closed form
 using elliptic integrals~\cite{2023PhRvD.108b4042G}. 

Setting $e=0$ in the expressions above trivially returns the equations used in Refs.~\cite{2015PhRvD..92f4016G, 2023PhRvD.108b4042G} for circular binaries where $r=a$. Moreover, it is immediate to prove that spin evolution of an eccentric binary on an orbit described by $a$ and $e$ is mathematically equivalent to that of a circular binary with orbital separation 
\begin{equation}
\label{trasf1}
a' = a \left(1-e^2\right)\,,
\end{equation}
provided one also changes the time variable to 
\begin{equation}
\label{trasf2}
t' = t \left(1-e^2\right)^{3/2}\,.
\end{equation}
This change of variables for space and time is reminiscent to the Lorentz transformations of special relativity, where the eccentricity $e$ enters the Lorentz factor $\gamma = 1/\sqrt{1-e^2}$ in place of the velocity $v/c$.

The spin properties of eccentric BH binaries can thus be mapped to those of an effective circular source. In other words, the eccentricity can be transformed away and dealt with in post-processing. We can thus make direct use of the numerical infrastructure that has already been developed for circular binaries: when tackling spin precession for eccentric sources, one can evolve the corresponding circular source and rescale the solution according to Eqs.~(\ref{trasf1}) and~(\ref{trasf2}).

\subsection{Radiation reaction} \label{sec:Radiation Reaction equations}

At lowest order, the eccentricity and semi-major axis evolve according to \citeauthor{1964PhRv..136.1224P}' equations \cite{1964PhRv..136.1224P}:
\begin{align} \label{eq:Peters1}
\frac{\dd a}{\dd t}&=-\frac{64}{5}\,\frac{q }{(1+q)^2}\,\frac{1+\frac{73}{24}e^2+\frac{37}{96}e^4}{a^3 \left(1-e^2\right)^{7/2}} \,M^3\,,\\
\label{eq:Peters2}
\frac{\dd e}{\dd t}&=-\frac{304}{15}\,\frac{q}{(1+q)^2}\frac{e \left(1+\frac{121}{304}e^2\right)}{a^4 \left(1-e^2\right)^{5/2}}\,M^3\,.
\end{align}
Equations~(\ref{l2}), (\ref{eq:Peters1}), and (\ref{eq:Peters2}) can be used to derive the radiation-reaction timescale: 
\begin{equation}
\label{taurrwithL}
\tau_{\rm rr} = \frac{ L}{|{\dd L}/{\dd t}|} = \frac{5}{32} \frac{(1+q)^2}{q M^3} 
\frac{ a^4 \left(1-e^2\right)^{5/2}
   }{1+\frac{7}{8}
   e^2}\,.
\end{equation}
Compared to other possible definitions including the more common $a/ |\dd a/\dd t|$, defining $\tau_{\rm rr}$ in terms of $L$ as in Eq.~(\ref{taurrwithL}) is more appropriate in our context because the derivative $\dd L/dt$ directly enters Eq.~(\ref{eq:dldt}).

Equations~(\ref{eq:Peters1}) and (\ref{eq:Peters2}) lead to several predictions 
when integrated either forward or backward in time from some initial condition ($a_0$, $e_0$). First, one has $\dd a /\dd t < 0$ for all values of $a$ and $e$, which is trivially related to the fact that GWs can only dissipate energy and not inject it into the system. This implies one can use $a$ as a time coordinate and consider $\dd e/ \dd a = \dd e / \dd t \times (\dd a/\dd t)^{-1}$. If $e_0=0$, one has $\dd e/\dd a =0 $ and the orbit remains circular throughout its evolution. If $e_0>0$, one has $\dd a / \dd e >0$, i.e. the eccentricity tends to decrease along with the semi-major axis, eventually approaching zero. Conversely, the eccentricity tends to increase when integrating backward in time up to $e=1$. 
The $e_0=0$ case is akin to a point of unstable equilibrium: while perfectly circular binaries stay circular even at past time infinity, binaries that are eccentric (even if arbitrarily close to circular!) become parabolic 
when traced back far enough in time.
In particular, for eccentric binaries one has \cite{1964PhRv..136.1224P}
\begin{equation}\label{petersasy}
\lim_{t\to -\infty} a \propto \frac{1}{1 - e^2}\,,
\end{equation}
which implies that, although $a\to \infty$ %
and $1-e^2 \to 0$, their product stays constant. 
The key consequence is that the angular momentum $L$ of an eccentric source from Eq.~(\ref{l2}) asymptotes to a constant value when backpropagated to past time infinity. This is in contrast with the circular case where instead the angular momentum $L\propto \sqrt{a}$ diverges for wide orbits. 

Such asymptotic distinction between eccentric and circular binaries plays a pivotal role in this study and constitutes a fundamental aspect of the interplay between spins and eccentricity in BH binaries.
The analytic results presented in Refs.~\cite{2015PhRvD..92f4016G, 2023PhRvD.108b4042G} in the $a \to \infty$ limit intrinsically rely on the divergence of $L$ and thus cannot be generalized to eccentric orbits, with conceptual consequences for GW population analyses \cite{2022PhRvD.105b4076M} that still need to be explored. We note, however, that the \citeauthor{1964PhRv..136.1224P}' equations themselves lose validity in this limit \cite{2019CQGra..36r5003M}. Equations~(\ref{eq:Peters1}) and (\ref{eq:Peters2}) have been orbit-averaged~\cite{1964PhRv..136.1224P} and thus are valid only when $\tau_{\rm orb}\ll \tau_{\rm rr}$. From Eqs.~(\ref{eq:tau_orb}), (\ref{taurrwithL}), and (\ref{petersasy}), it is straightforward to prove that the two timescales are of the same order of magnitude when $a\to \infty$ and $e\to 1$ (cf. Sec.~\ref{sec: Timescales}). %

\subsection{Precession-averaged inspiral}

We now proceed on extending the precession-average formalism to eccentric binaries. A generic quantity $X$ can be averaged to
$\langle X \rangle = \left(\int_{0}^{\tau_{\rm pre}} X \dd t \right) / \tau_{\rm pre}$, where the integral is more conveniently implemented using Eq.~(\ref{eq:ddeltachidt}); cf. Ref.~\cite{2023PhRvD.108b4042G}.
Once constant of motions are taken into account and both the orbital and the precessional motion are averaged over, one only needs to connect quantities that vary on radiation-reaction timescale. 

For the circular problem, there are only two of such quantities, namely the magnitudes of the orbital and total angular momenta. The evolution thus reduces to integration a single ordinary differential equation (ODE)~\cite{2015PhRvL.114h1103K,2015PhRvD..92f4016G}: 

\begin{equation}\label{eq:dkdu}
    \frac{\dd \kappa}{\dd u}=\langle |\bm{S}_1+\bm{S}_2|^2 \rangle,
\end{equation}
where 
\begin{equation}
u=\frac{1}{2L}
\end{equation}
and 
\begin{equation}
\kappa=\frac{J^2-L^2}{2L}
\end{equation}
are convenient reparametrizations of $L$ and $J$, respectively. The right-hand side of Eq.~(\ref{eq:dkdu}) depends on quantities that are either constants of motion or vary on $\tau_{\rm rr}$ and can be written down in closed form~\cite{2021arXiv210610291K, 2023PhRvD.108b4042G}.
The key assumptions used in the derivation of Eq.~(\ref{eq:dkdu}) is that $\dd \boldsymbol{L} / \dd t$ is parallel to $\boldsymbol{L}$ and independent of the spins~
\cite{2015PhRvL.114h1103K,2015PhRvD..92f4016G}, which are still true for eccentric binaries at leading order in radiation reaction \cite{2004PhRvD..70l4020S}. We can thus use Eq.~(\ref{eq:dkdu}) as it is, provided one properly generalizes $L$ as in Eq.~(\ref{l2}). 

The eccentric problem however has three, and not two, quantities that vary on the radiation reaction time: $a$, $e$, and $J$ (or some equivalent reparametrization). We already have two variables, $\kappa$ and $u$. For the third, we choose
\begin{equation}
u_{c} \equiv u(a, e=0) = \frac{(1 + q)^2}{2 q M^2}\sqrt{\frac{M}{a}}\,,
\end{equation}
which is the coordinate used previously for the integration of circular binaries and does not depend on $e$. In particular, the condition $0\leq e<1$ implies $u\geq u_c>0$. 
The resulting evolutionary equation can be easily derived from 
Eq.~(\ref{eq:Peters1}) and (\ref{eq:Peters2}) and reads:
\begin{equation}
\label{odeuuc}
 \frac{\dd u}{\dd u_c} = -\frac{12 u_c u \left(7 u_c^2-15
   u^2\right)}{37 u_c^4-366 u_c^2 u^2+425
   u^4}\,.
   \end{equation}
With some tedious but straightforward algebraic manipulation, the solution can be written down as 
      \begin{equation}
      \label{implicit}
u_c u^{37/84} \left(\frac{u^2}{u_c^2}-1\right)^{121/532}
   \left(\frac{u^2}{u_c^2}-\frac{121}{425}\right)^{145/532} = k_0\,,
   \end{equation}
where $k_0$ is a constant that needs to be determined from the initial conditions $u_{\rm c0}$ and $u_0$. 

 For an initially circular binary, one has $u_0= u_{c0}$ and $k_0=0$. The only solution to Eq.~(\ref{implicit}) is indeed $u=u_c$, i.e. circular binaries stay circular. For an initially eccentric binary, one has $k_0>0$ and Eq.~(\ref{implicit}) can be solved numerically to identify $u(u_c)$. This is simpler and more accurate than numerically integrating Eq.~(\ref{odeuuc}). Compared to the analytic expression for $a(e)$ reported in Ref.~\cite{1964PhRv..136.1224P}, our formulation has the key advantage of being regular in the limit of circular binaries ---the first term in parenthesis in Eq.~(\ref{implicit}) acting as a regularizer--- and thus more amenable to numerical evaluations.

In summary, given a set of constants of motion ($q$, $\chi_1$, $\chi_2$, $ \chi_{\rm eff}$) and initial conditions ($u_{c0}$, $u_0$, $\kappa_0$), performing a precession-averaged evolution of an eccentric binary reduces to integrating the ODE (\ref{eq:dkdu}) under the constraint imposed by Eq.~(\ref{implicit}). The ODE solver is identical to that of the circular case and the root finder for $u(u_c)$ is an additional, but trivial, computational task. Overall the performance of our implementation is similar to what we reported in Ref.~\cite{2023PhRvD.108b4042G} for the circular case.

\subsection{Timescale separation}\label{sec: Timescales}
\begin{figure}[t]\label{fig:timescale_window}
    \includegraphics[width=\columnwidth]{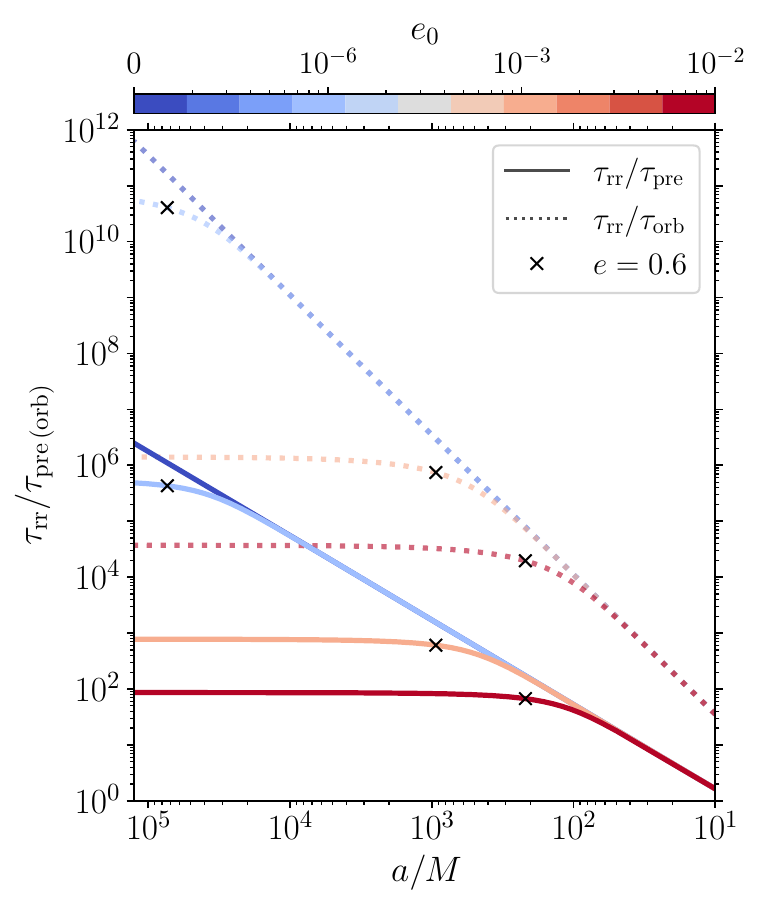} 
    \caption{Evolution of the timescale ratios $\tau_{\rm rr}/\tau_{\rm pre}$ (solid) and $\tau_{\rm rr}/\tau_{\rm orb}$ (dotted) as a function of the binary semi-major axis for four GW sources with $q=0.5$, $\chi_1=0.6$, $\chi_2=0.9$, and $(\theta_1,\theta_2,\Delta\Phi)=(\pi /3,\pi/ 4,-\pi/ 3)$. %
These are initialized with different eccentricities $e_0 = 0$, $10^{-6}$, $10^{-3}$, $10^{-2}$ at a separation $a =10$ M as indicated on the color scale and propagated backward. Crosses indicate the location of the inspiral where the eccentricity reaches $e=0.6$.}

  \end{figure}

We now investigate the validity of our formalism. Averaging over the orbital and precessional motions in sequences relies on the timescale separation $\tau_{\rm orb} \ll \tau_{\rm pre} \ll \tau_{\rm rr}$. In the circular limit, this inequality trivially correspond to the PN condition $a=r\gg GM/c^2$. 
As illustrated in Fig.~\ref{fig:timescale_window}, %
 the eccentric case is less straightforward.

 In particular, one has

\begin{equation} 
\label{eq:lim_rr/orb}
 \frac{\tau_{\rm rr}}{\tau_{\rm orb}} \propto
    \begin{cases}
        a^{5/2}  & \qquad  {\rm if }\quad  e\to 0\,,
        \\
      \rm const.  & \qquad  {\rm if }\quad e\to 1 \,,
\end{cases}
\end{equation}
and 
\begin{equation}
\label{eq:lim_rr/pre}
\frac{ \tau_{\rm rr}}{\tau_{\rm pre}} \propto
    \begin{cases}
    a^{3/2}  &  \qquad  {\rm if }\quad  e\to 0\,,
    \\
        \rm const. & \qquad  {\rm if }\quad e\to 1 \,,
\end{cases}
\end{equation}
where the circular (parabolic) limit corresponds to the late time (early time) behavior of a generic eccentric source. It follows that, for eccentric binaries, the accuracy of the multitimescale approach does not keep on improving as one moves to larger and larger separations. Rather, it plateaus. From Fig.~\ref{fig:timescale_window}, the transition between the two behaviors takes place when $e \sim 0.5$.

This is qualitatively different than the circular case, where multi-timescale evolutions are inaccurate close to merger (because of the breakdown of the PN approximation) but become increasingly accurate toward past time infinity \cite{2015PhRvD..92f4016G}.

\begin{figure}[!ht] \label{fig:timescale_map}
\includegraphics[width=\columnwidth]{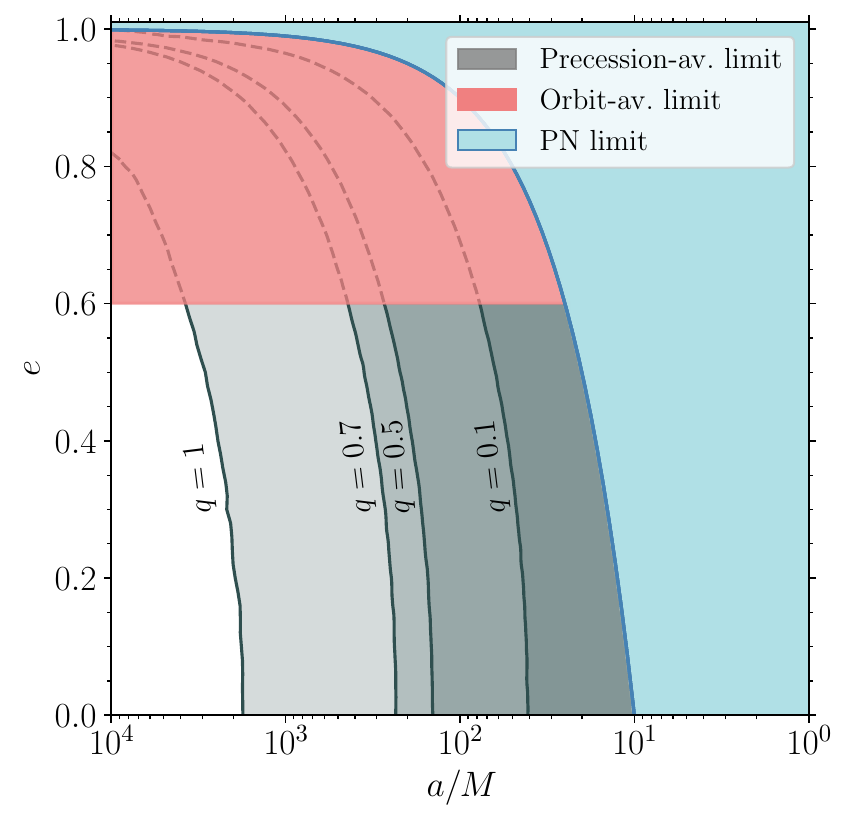} 
   \caption{Validity region of the precession-average formalism as a function of binary semi-major axis $a$ and eccentricity $e$.
The light blue area indicates the region where the close-passage distance is $a(1-e)< 10 M$ , likely beyond the regime of validity of the PN approximation. The red area indicates the condition $e>0.6$, which is our conservative limit for the inapplicability of the orbit-averaged approach to radiation reaction. The black solid curves show the condition $\tau_{\rm rr}/\tau_{\rm pre}=100$, with $>100$ being true in the lighter regions to the left of each curve. %
We average the  value of $\tau_{\rm rr}/\tau_{\rm pre}$ over four sets of binaries %
with fixed mass ratios $q=0.1,0.5,0.7,1$ and spins that are generated uniformly in magnitudes in $[0.1,1]$ and isotropic in directions. %
}
\end{figure}

We express the regime of validity of the multi-timescale approach in terms of two approximate conditions:
\begin{itemize}
\item[(i)] 
At small separations, the periapsis distance should be sufficiently large for the equations to remain PN-valid. Borrowing the threshold from Ref.~\cite{2009PhRvD..80h4043B}, we set $a(1-e) \gtrsim 10 M$.
\item[(ii)] 
At large separations, the \citeauthor{1964PhRv..136.1224P}' equations predict parabolic orbits. These have infinitely large periods, which is manifestly against the quasi-adiabatic approach used to derive the equations themselves. This is a known issue \cite{2017CQGra..34m5011L} and, in particular, Refs.~\cite{2019CQGra..36b5004L, 2019CQGra..36r5003M} suggest that orbit averaging can only be meaningfully applied if $e\lesssim 0.6$.
\end{itemize}

These forbidden regions are highlighted in Fig.~\ref{fig:timescale_map}. In the remaining grey-shaded region, the applicability of our formalism depends on the precession timescale. Figure~\ref{fig:timescale_map} shows contours of $\tau_{\rm pre}/\tau_{\rm rr}$ for populations of BH binaries with different values of the mass ratio, averaging over both spin magnitudes and directions.
 Regions to the left of each curve correspond to $\tau_{\rm pre}/\tau_{\rm rr}>100$, which is safe territory for the multi-timescale approach. This is not dissimilar to the circular case \cite{2015PhRvD..92f4016G} and, indeed, the black curves in Fig.~\ref{fig:timescale_map} remain largely vertical, i.e. independent of the eccentricity. The $e=0.6$ limit indicated in condition (ii) above roughly separates the region of the parameter space where the $\tau_{\rm pre}/\tau_{\rm rr}$ contours start bending leftward. 

For a quick rule of thumb, our multi-timescale approach  can be applied to eccentric binaries much like in the circular case as long as the eccentricity is small to moderate, with $e = 0.6$ providing a nominal threshold. The high-eccentricity case requires a new formalism, perhaps averaging over the various phenomena entering the dynamics (orbit, precession, inspiral, eccentricity decay) in a different order.

\section{Phenomenology}\label{appsection}  %

{We first describe our findings in terms of the parameter that most directly enters our formalism. We then present predictions for quantities that are more directly observable.}

\subsection{Weighted spin difference} \label{sec: Spin-eccentricity interplay}

Compared to the quasi-circular case, eccentricity enters our formalism in two key aspects. First, the coordinate transformation of Eqs.~(\ref{trasf1}) and (\ref{trasf2}) accelerates the evolution of all quantities that vary on the precession timescale, including the spin directions (this feature was already noted in Ref.~\cite{1998PhRvD..58l4001G}). Second, the orbital angular momentum $L$ does not diverge as $t \to -\infty$ \cite{1964PhRv..136.1224P}. 

For quasi-circular binaries, the right-hand side of Eq.~(\ref{eq:ddeltachidt}) tends to zero in this limit, which implies that at large separations BH spins move on precession cones of constant opening angles $\theta_{1,2}$ \cite{2023PhRvD.108b4042G}. This is not true in general for eccentric binaries, %
 suggesting that, even at early times, the BH spin continue to nutate with an amplitude that approaches a constant value and does not degenerate to zero (but see Sec.~\ref{sec: Timescales} on the validity of our formalism).%

\begin{figure}[t]\label{fig:deltachi_time}
    \includegraphics[width=\columnwidth]{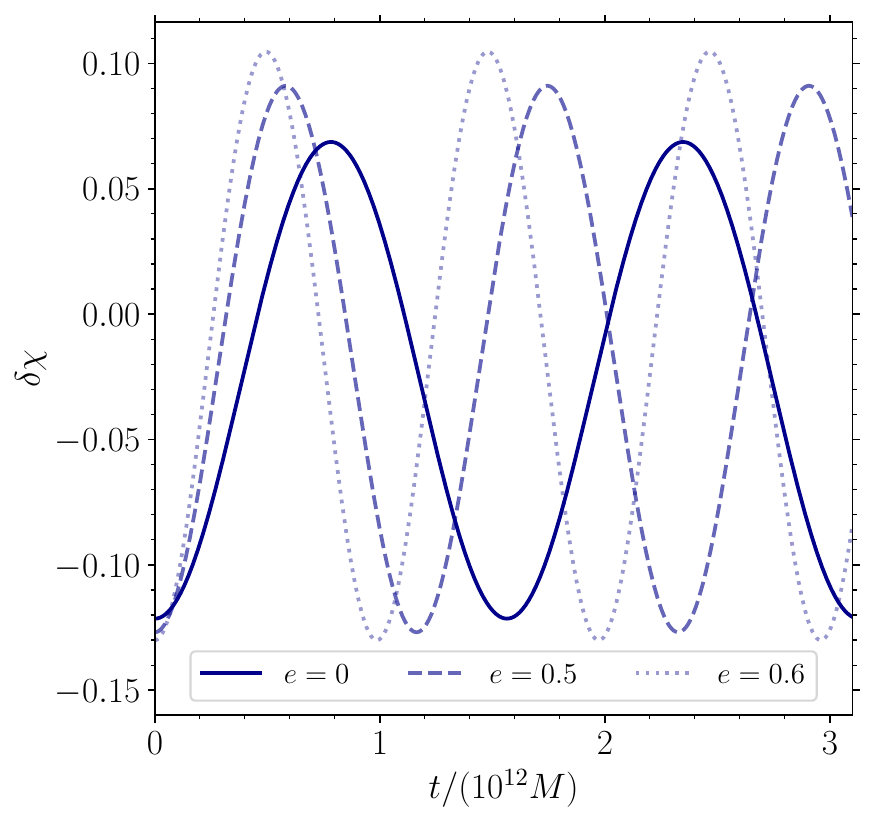} 

    \caption{
    Time evolution of the weighted spin difference $\delta \chi$ for three representative BH binaries with  mass ratio $q=0.95$, spin magnitudes {$\chi_{1}=\chi_2=0.9$}, initial spin orientations $(\theta_1,\theta_2,\Delta\Phi)=(\pi /3,\pi/ 4,-\pi/ 3)$, semi-major axis $a=10^4 M$, and eccentricities $e=0$ (solid), 0.5 (dashed), and 0.6 (dotted). %
 In terms of the deviation parameter from Eq.~(\ref{eq:Delta}), these three sources have $\Delta = 0$, $1.5 \times 10^{-3}$, and $2.4 \times 10^{-3}$, respectively.%
 }%

\end{figure}

Figure~\ref{fig:deltachi_time} shows the precession-timescale evolution  (i.e. without radiation reaction)  of the weighted spin difference $\delta \chi$ for three binaries %
 with different eccentricities.
Notably, the spins of eccentric binaries oscillate with a shorter period and a larger amplitude. These are a direct consequences of the two dynamical features we just highlighted. %

\begin{figure*}[t]\label{fig:griglia}
    \includegraphics[width=0.98\textwidth]{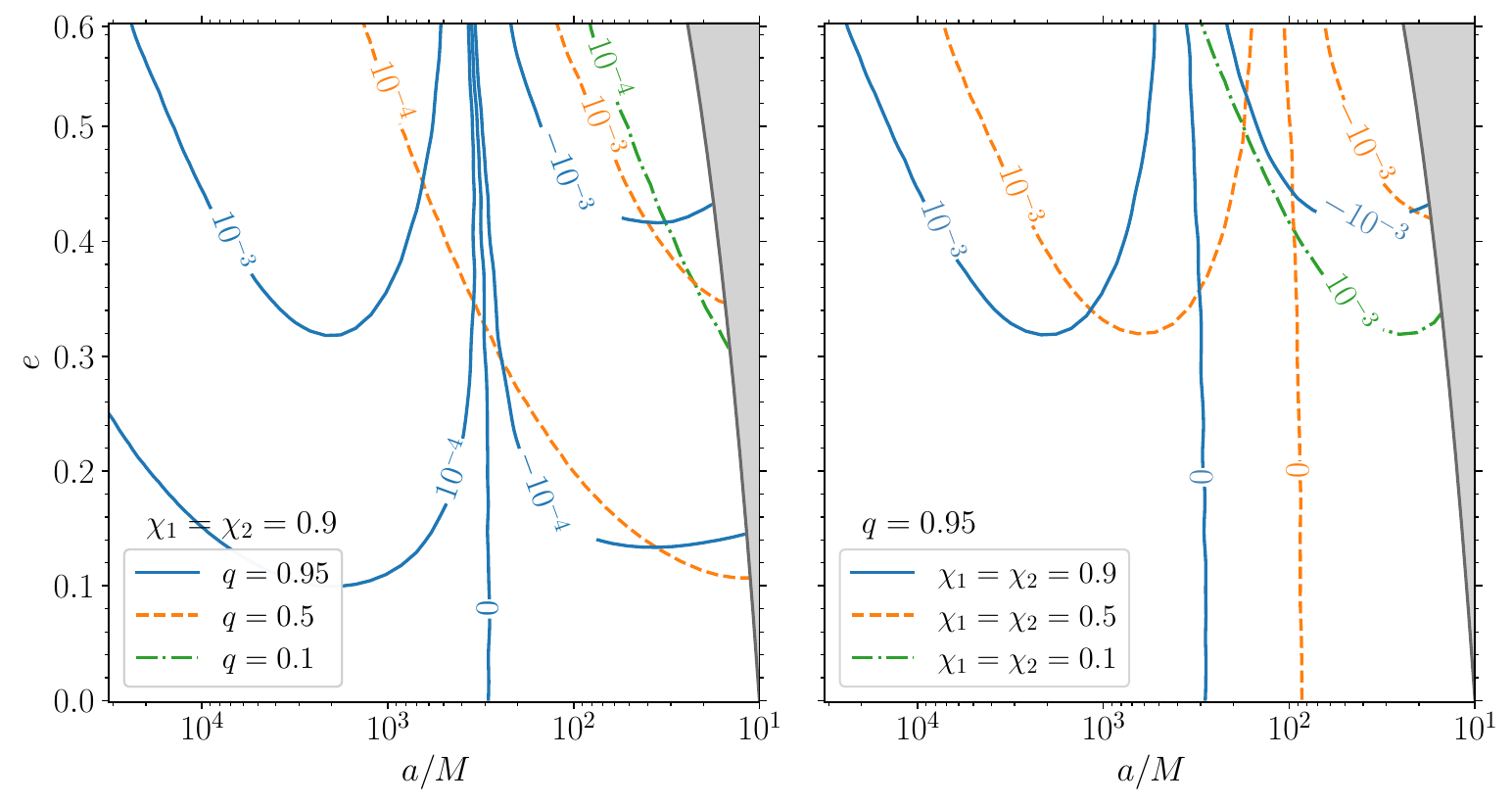} 
    \caption{
Spin-eccentricity interplay in terms of the deviation parameter $\Delta $ as a function of binary semi-major axis $a$ and eccentricity $e$. We consider sources with $e<0.6$, %
which sets the validity of our model (Sec.~\ref{sec: Timescales}). 
For each panel, we consider three sets of  BH binaries with different mass ratios and spins magnitudes. In all cases, we average over the orientations of the spins, which are assumed to be  distributed  isotropically.
In the right panel, we set {$\chi_{1}= \chi_{2}=0.9$} and vary $q=0.95$ (blue solid), $0.5$ (orange dashed), and $0.1$ (green dotted). %
In the left panel, we instead set $q=0.95$ and vary {$\chi_{1}= \chi_{2}=0.9$} (blue solid), $0.5$ (orange dashed), and $0.1$ (green dotted). The blue solid contours are the same across the two panels. {Grey areas indicate the region $a(1-e)\leq10M$ where the PN approximation breaks down.}  %
}
\end{figure*}

We further quantify the impact of eccentricity on spin precession with a suitable deviation parameter $\Delta$.  We first precession-average $\delta \chi$ and rescale it to its oscillation amplitude~\cite{2023PhRvD.108b4042G}
 \begin{align}\label{eq:deltachitilde}
\langle \delta \tilde{ \chi} \rangle&= \frac{ \langle \delta \chi \rangle - \delta \chi_-}{\delta\chi_+ - \delta\chi_-} \in [0,1]\,.
 \end{align}
We then compute $\langle \delta \tilde{ \chi} \rangle$ for a given eccentric binary and compare it against an equivalent quantity $\langle \delta \tilde{ \chi} \rangle_c$, which is estimated assuming the same values of $a$, $q$, $\chi_1$, $\chi_1$, $\theta_1$, $\theta_2$ and $\Delta \Phi$ but setting  $e=0$. Our deviation parameter is then defined as
 \begin{align}\label{eq:Delta}
\Delta &= \frac{\langle \delta \tilde{ \chi} \rangle - \langle \delta \tilde{\chi} \rangle_c}{\langle\delta \tilde{ \chi} \rangle+\langle \delta \tilde{ \chi} \rangle_c} \in [-1,1]\,,
 \end{align} 
and can be interpreted as a fractional measurement of the impact of eccentricity on spin precession.
Figure~\ref{fig:griglia} illustrates the value of $\Delta$ across some sections of our parameter space.
We consider six sets with fixed mass ratios %
and spin magnitudes %
and average our results over spins directions distributed isotropically.

 In general, we find that the magnitude $|\Delta|$ %
 increases with both the mass ratio and the spin magnitudes, %
signaling an enhanced interplay between spins and eccentricity in such systems. %
The deviation $\Delta$ is largely positive across the parameter space, with the key exception of binaries with nearly equal masses and large spins. %
Note that $\Delta$ is a precession-avarage quantity and as such is only accurate when the underlying timescale hierarchy is respected. %

 \subsection{Spin orientations} \label{sec: Consequence on spin orientations}

\begin{figure*}[p]
    \includegraphics[width=\textwidth]{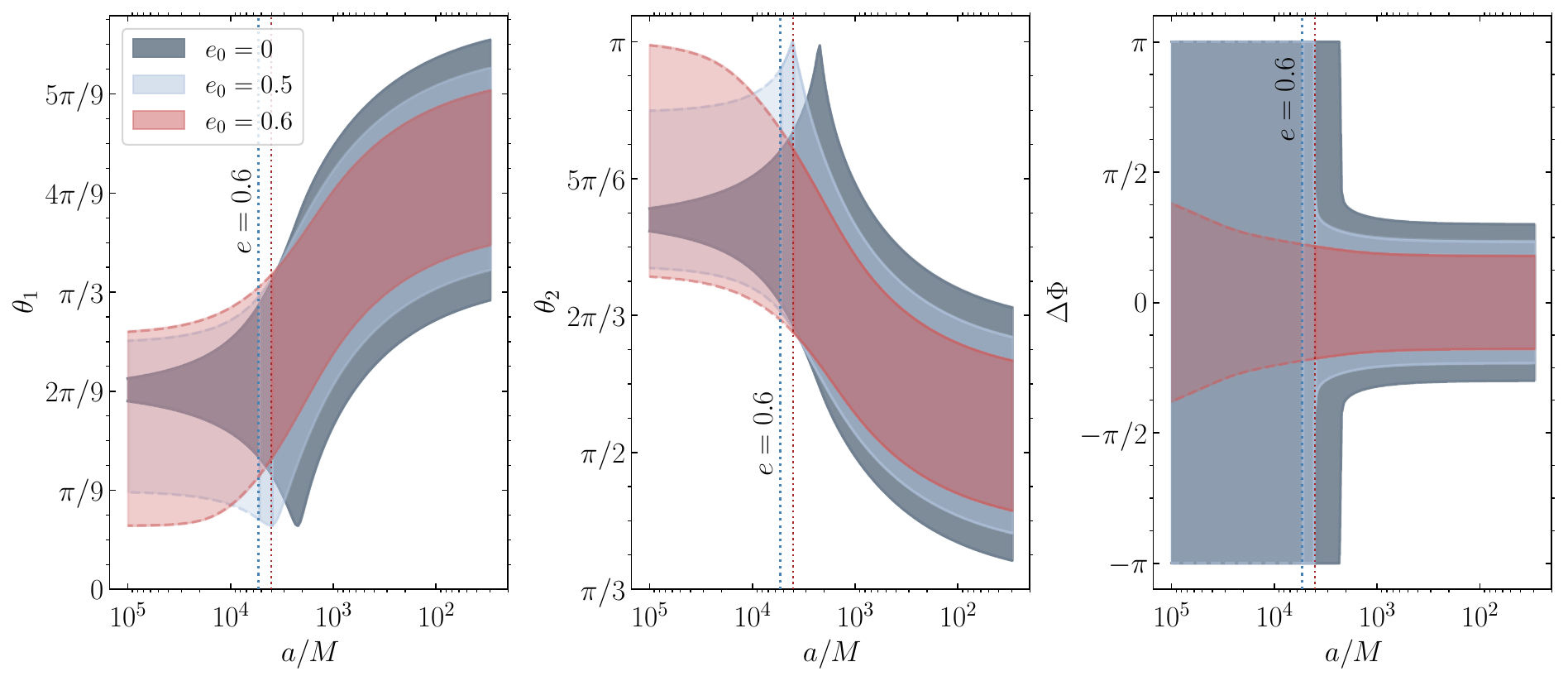} 

    \caption{
    Evolution the spin angles $\theta_1$ (left), $\theta_2$ (middle), and $\Delta \Phi$ (right) under radiation reaction. Each of the angles oscillate in the reported ranges. We consider three binaries with mass ratio $q=0.95$, spin magnitudes {$\chi_1=\chi_2=0.96$},  and initial spins orientation $(\theta_1, \theta_2, \Delta \Phi)=(\pi/3,2 \pi/3,-\pi/9)$. Sources are evolved assuming  initial eccentricities $e_0=0$ (dark grey), $0.5$ (light grey), and $0.6$ (red) at $a_0= 4000 M$. We track the evolution both forward to $a=30 M$ and backward to $a=10^5 M$. At large separations, the eccentricity grows beyond $e=0.6$ (vertical dotted lines) and our formalisms loses validity (lighter shaded areas and dashed curves).}%

    \label{fig:trio}
\bigskip   \bigskip\bigskip 
    \includegraphics[width=\textwidth]{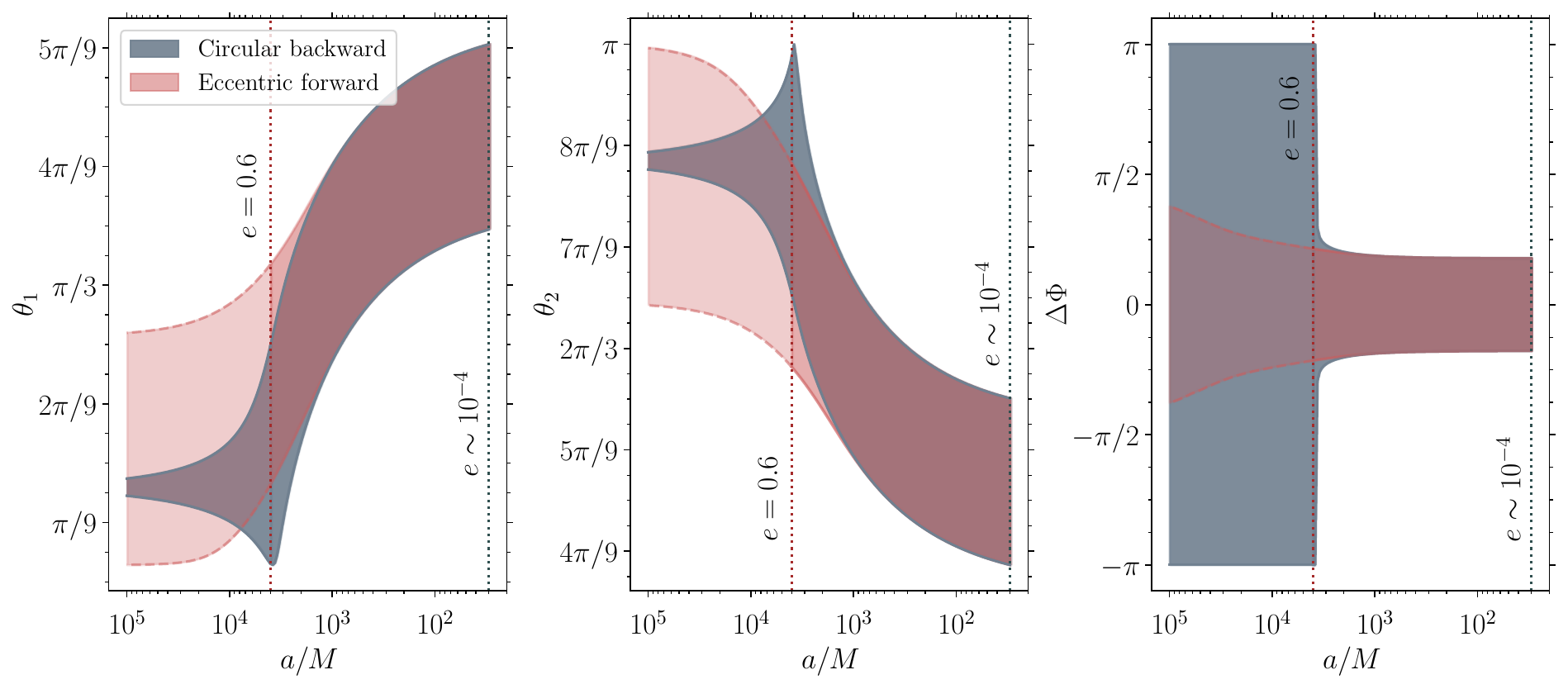} 
    \caption{
    Forward and backward evolution of the spin angles $\theta_1$ (left), $\theta_2$ (middle), and $\Delta \Phi$ (right), resetting the eccentricity to zero near merger. We consider the same binary of Fig. \ref{fig:trio} with $q=0.95$,  {$\chi_1=\chi_2=0.96$}, $(\theta_1, \theta_2, \Delta \Phi)=(\pi/3,2 \pi/3,-\pi/9)$, $a_0= 4000 M$, and $e_0=0.6$, and evolve it toward merger (red). At $a=30 M$, this binary has an eccentricity $e\sim 10^{-4}$. We artificially set this value to zero and evolve the same binary backward to large separations (dark grey). %
 } 
 \label{fig:BF}
   \end{figure*}

While $\delta\chi$ is the quantity that directly enter our formalism, more intuitive insight %
 can be gained by rephrasing our findings in terms of the spin angles $\theta_1$, $\theta_2$, and $\Delta \Phi$. Our results are shown in Figs.~\ref{fig:trio} and \ref{fig:BF} for some representative sources. Note we only illustrate the range of variation of each of these angles as the timescale separation $\tau_{\rm pre} \ll \tau_{\rm rr}$ implies they oscillate many times at each orbital separation.

Figure~\ref{fig:trio} shows the evolution of three systems with $q=0.95$, {$\chi_{1}=
\chi_2=0.96$}, and initial spins orientation $(\theta_1, \theta_2, \Delta \Phi)=(\pi/3,2 \pi/3,-\pi/9)$. Integrations are initialized at the initial semi-major axis $a_0=4000 M$ considering three different initial eccentricities $e_0=0$, $0.5$, and $0.6$. While we can formally integrate both forward and backward in time (i.e. toward smaller and larger eccentricities, respectively), we expect our formalism to break down at early times when $e\gtrsim 0.6$, cf. Sec.~\ref{sec: Timescales}.
The key message here is that the spins of these binaries 
trace different precession cones as they evolve toward merger. That is, the spin dynamics depends on the eccentricity. 

As these binaries approach merger, the range the spin angles can vary within is smaller for sources with larger initial eccentricities. It is worth noting that this is true even if the eccentricity of all three sources close to merger is essentially zero ($e\sim 10^{-4}$ at $a=30M$): the spins of BH binaries at small separations (i.e. when they become detectable by our instruments) ``remember'' their past evolution on eccentric orbits.
The smaller nutation amplitude observed close to merger for the eccentric evolutions compared to the circular case is consistent with the results reported in Sec.~\ref{sec: Spin-eccentricity interplay} for BH binaries with nearly equal masses and high spins. While this is the case for the sources in Fig.~\ref{fig:trio}, we do not expect this to be a generic feature. %

The strong features observed at { $a\simeq 2210 M $ ($4000 M$)} %
and the azimuthal angle $\Delta\Phi$ is instantaneously ill-defined. %
The occurrence of these phase transitions is deeply affected by the eccentricity to the point that, at least for these specific cases, the transition itself disappears in the case of the most eccentric source. This point is further explored in Sec.~\ref{sec: Spin precession morphologies} below.

While likely outside the regime of validity of our formalism, the backward integration of these sources toward larger separation shows how eccentric sources maintain a finite nutation amplitude, i.e. the range of variations of $\theta_{1,2}$ does not go to zero as $a$ increases. This is a direct consequence of the magnitude of the orbital angular momentum remaining finite at past time infinity. %

The eccentric binaries of Fig.~\ref{fig:trio} reach merger with eccentricities of $\ssim 10^{-4}$, which is well below the distinguishability threshold of our detectors~\cite{2018PhRvD..98h3028L}. In practice, these sources will all be classified as circular. %
Figure~\ref{fig:BF} demonstrates the impact of such indistinguishability on the reconstruction of the spin history. 
We consider the same $e_0=0.6$ evolution of Fig.~\ref{fig:trio}, where the source is evolved from large to small separations. We then take the final condition, reset its residual eccentricity to $0$, and evolve it back to the initial large separation. This ``back and forth'' evolution is \emph{not} invertible, i.e. the binary does not go back to where it started in parameter space. %

As shown in Fig.~\ref{fig:BF}, the two evolutions are similar to each other only at small separations when the eccentricity is low and differ  as $e$ increases. For context, the forward evolution had $e\sim 0.14$ at $a=1000M$. The backward evolution (i.e. which  in this example corresponds to our reconstructed information) shows a prominent phase transition that is instead absent in the eccentric forward evolution (i.e. which instead corresponds to the actual source). %
This experiment shows that eccentricity, even when it is far below the threshold of distinguishability, poses a serious challenge when attempting a reconstruction of the binary formation history using the spin directions~\cite{2016ApJ...832L...2R,2017Natur.548..426F,2017NatCo...814906S,2021NatAs...5..749G,2013PhRvD..87j4028G,2018PhRvD..98h4036G,2021PhRvD.103f3032S}. Residual eccentricity is a systematic uncertainty that needs to be taken into account when interpreting GW~data. %

 \subsection{Spin morphologies}\label{sec: Spin precession morphologies}

The parameter space of precessing BH binaries can be divided into three distinct regions, or ``morphologies,'' according to the evolution of the azimuthal angle $\Delta \Phi$~\cite{2015PhRvD..92f4016G}. In particular, this angle can either (i) circulate between $\Delta \Phi=0$ to $\Delta \Phi=\pm\pi$, (ii) librate about $\Delta \Phi=0$ (and never reach $ \Delta \Phi=\pm \pi$), or (iii) librate about $\Delta \Phi=\pm \pi$ (and never reach $\Delta \Phi=0$). In particular, the zero-amplitude limits of the two librating morphology are the so-called spin-orbit resonances of the spin precession problem~\cite{2004PhRvD..70l4020S}. 

While the morphology is an integrated quantity that is defined over an entire precession cycle, radiation reaction causes secular transitions between the different classes.  %
Some examples are shown in Figs.~\ref{fig:trio} and \ref{fig:BF}, where in both cases the $e=0$ and $e=0.5$ binaries belong to the circulating morphology at large separations and transition toward the librating about $\Delta\Phi=0$ morphology at some point during the inspiral. Classifying BH binaries in terms of these classes is promising because, at least for the quasi-circular problem, the morphology near merger (where binaries are detected) tracks the spin tilts at large separations (where binaries are formed) \cite{2015PhRvD..92f4016G,2020CQGra..37v5005R}.

For circular sources, one can prove that all binaries belong to the circulating morphology at large separations, while the two librating morphologies can only be populated by transitions occurring during the inspiral \cite{2023PhRvD.108b4042G}. Once more, this feature is due to the divergence of $L$ at early times and does not hold for eccentric sources. This implies that eccentric binaries will more likely be found in the two librating morphologies compared to their circular counterparts. 

\begin{figure}\label{fig:morph}
    \includegraphics[width=\columnwidth]{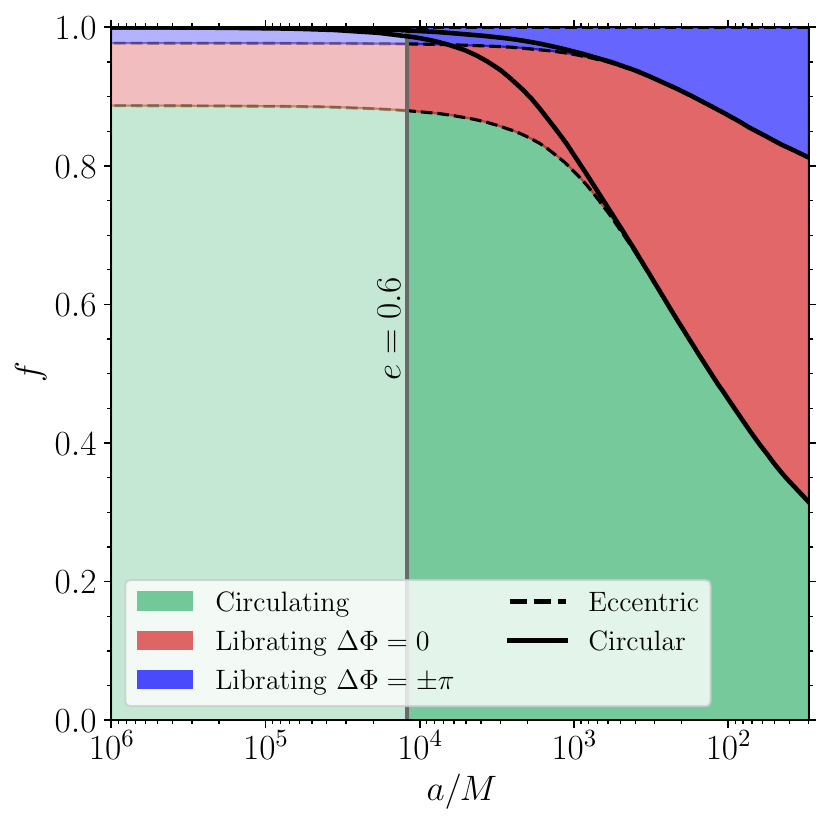} 
    \caption{ Fraction of eccentric (colored areas and dashed black curves) and circular (solid black curves) binaries falling in each of the three morphologies as a function of the binary semi-major axis $a$. %
   We assume a set of BH binaries with fixed mass ratio $q=0.95$, spin magnitudes $\chi_{1,2}$ uniformly distributed in $[0.5,1]$,  %
   and isotropic spin directions. Sources are initialized at $(a_0=30M, e_0=2\times10^{-3})$ and evolved backward to $a=10^{6}M$. Lighter areas to the left of the vertical grey line indicate the region where $e>0.6$ and our formalism loses validity.} 

\end{figure}

This point is illustrated in Fig.~\ref{fig:morph}. We consider BH binaries with $q=0.95$, $\chi_{1,2}$ uniformly distributed in $[0.5,1]$, 
and spin orientations distributed isotropically. Sources are evolved backward in time from small ($a_0=30 M$) to large separations ($a_{\rm f}=10^{6} M$) with an initial eccentricity $e_0=2 \times 10^{-3}$. We report the fraction of binaries in each of the spin morphologies, together with analogous fractions obtained setting $e_0=0$. %
The key conclusion is that, as the evolution proceeds, some eccentric sources remain in their librating morphologies at large separations while all circular binaries transit to circulation.
This might have important consequences when attempting to use the spin morphologies to constrain astrophysical formation channels.

\section{Conclusions}\label{sec: Conclusion}

Spin precession and eccentricity are precious indicators of the astrophysical origin of binary BHs in both the stellar-mass and the supermassive regimes. As binaries evolve from formation to merger, couplings between the orbital eccentricity and the BH spins enters the dynamics, with potential consequences for our astrophysical inference. 

Binaries tend to circularize during their long inspiral before merger \cite{1964PhRv..136.1224P}, which implies most sources are expected to enter the sensitivity window of our detectors with vanishingly small eccentricities, likely below the distinguishability threshold \cite{2018PhRvD..98h3028L,2023arXiv230713367G}. This paper shows that 
 the spin evolution retains some memory of the eccentric past of BH binaries. The implications are twofold, with a pro and a con:
 \begin{itemize}
 
 \item[(i)] Because the spin orientations depend on the eccentricity, interplay between the two effects provides an avenue to infer that sources \emph{formed} on eccentric orbits even if they are not \emph{detected} as such. Astrophysical degeneracies will likely prevent us from inferring this effect on an event-by-event basis, but the implications for the GW population problem are promising. If a sizable fraction of BH binaries forms on eccentric orbits, this should have an impact on the statistical properties of the spin orientations inferred via GWs. The next %
  step is to formalize this intuition in terms of concrete observational predictions. 
 
\item[(ii)] The interplay between orbital eccentricity and spin precession implies that inference using the latter will be polluted by the former. The broader program of inferring the astrophysical formation channels of BH binaries using the spin directions thus suffers from an important systematics, namely residual eccentricity. Analyses that attempt reconstructing the spin history of BH binaries should be carried out taking residual eccentricity into account. The next step here is to design a suitable strategy to ``marginalize out'' this source of uncertainty.

 \end{itemize}

Our paper presents a multi-timescale treatment of the binary dynamics able to capture moderately eccentric sources, generalizing results that were previously restricted to circular orbits~\cite{2015PhRvD..92f4016G,2023PhRvD.108b4042G} (see also Refs.~\cite{2019PhRvD.100l4008P,2020PhRvD.102l3009Y,2021arXiv210610291K,2022PhRvD.106b3001J}). %
 This extension involves two key steps: an analytical rescaling of the spin-precession equations and an additional prescription for the evolution of the eccentricity itself. We explored the resulting phenomenology showing in particular how eccentricity impacts (i)~the dynamical quantities entering our formalism, (ii)~the nutation amplitude of the BH spins and (iii)~the so-called spin morphologies. %

 Our findings will  be implemented in the public version of the \textsc{precession} code \cite{2016PhRvD..93l4066G,2023PhRvD.108b4042G}. It is important to stress that our approach relies on the well-known, orbit-averaged PN equations of motion \cite{1964PhRv..136.1224P}, which cannot be used reliably when the eccentricity is $e\gtrsim 0.6$ \cite{2019CQGra..36r5003M}. Pushing our formalism beyond this limit is a promising avenue for future work, which likely requires going back to the drawing board and carefully consider how the different timescales separate across the parameters space.

\acknowledgements

We thank Nick Loutrel, Isobel Romero-Shaw, Ulrich Sperhake, and Matt Mould for discussions.
G.F. and D.G. are supported by ERC Starting Grant No.~945155--GWmining, 
Cariplo Foundation Grant No.~2021-0555, MUR PRIN Grant No.~2022-Z9X4XS, 
and the ICSC National Research Centre funded by NextGenerationEU. 
G.F. is supported by Sigma Xi Grant No. G20230315-4609 and an Erasmus+ scholarship.
D.G. is supported by Leverhulme Trust Grant No.~RPG-2019-350 and MSCA Fellowship No.~101064542--StochRewind.
Computational work was performed at CINECA with allocations 
through INFN and Bicocca.

\bibliography{preccentric}

\end{document}